\documentstyle[prl,aps,epsf,multicol]{revtex}
\begin{document}
\title{\bf Survival of a Diffusing Particle in a Transverse Shear Flow: 
A First-Passage Problem with Continuously Varying Persistence Exponent}
\author{Alan J. Bray and Panos Gonos}
\address{Department of Physics and Astronomy, University of 
Manchester, Manchester M13 9PL, UK}

\date{\today}

\maketitle

\begin{abstract}
We   consider    a   particle   diffusing    in   the   $y$-direction,
$dy/dt=\eta(t)$,   subject  to   a  transverse   shear  flow   in  the
$x$-direction, $dx/dt=f(y)$, where $x \ge 0$ and $x=0$ is an absorbing
boundary.  We  treat  the class  of  models  defined  by $f(y)  =  \pm
v_{\pm}(\pm y)^\alpha$  where the upper  (lower) sign refers  to $y>0$
($y<0$).  We show that  the particle  survives with  probability $Q(t)
\sim  t^{-\theta}$ with $\theta  = 1/4$,  independent of  $\alpha$, if
$v_+=v_-$.  If  $v_+  \ne  v_-$,  however, we show that $\theta$  
depends  on  both $\alpha$ and the ratio $v_+/v_-$, and we determine 
this dependence.

\noindent

\end{abstract}

\begin{multicols}{2}

First-passage  problems  are an  important  aspect  of  the theory  of
stochastic  processes.    Applications  include  population  dynamics,
chemical reactions,  and many other problems  in science \cite{Guide}.
This class of  problems has attracted a resurgence  of interest in the
last decade.   The number of exactly-solved models,  however, is still
quite  small \cite{Guide,majumdar_review}.  The  classic example  is a
one-dimensional random walker, $dx/dt = \eta(t)$, where $\eta(t)$ is a
Gaussian  white   noise  with   mean  zero  and   correlator  $\langle
\eta(t)\eta(t')\rangle  = 2D\delta(t-t')$. If  the particle  starts at
$x>0$,  and there  is an  absorbing  boundary at  $x=0$, the  particle
survives   at    time   $t$   with   probability    $Q(x,t)   =   {\rm
erf}(x/\sqrt{4Dt})$  \cite{Guide}. For  large $t$  this  gives $Q(x,t)
\sim x/(Dt)^{1/2}$, defining an  exponent $\theta = 1/2$ (often called
a `persistence exponent'  \cite{majumdar_review}) through the relation
$Q(x,t) \sim t^{-\theta}$ for large $t$.

In this  paper we introduce a  new class of models  for which $\theta$
can be determined exactly and has, in general, a nontrivial value.  We
treat a  simple model  of a particle  diffusing in  the $y$-direction,
$dy/dt = \eta(t)$, but subject to a transverse flow field (`shear') in
the  $x$-direction,  $dx/dt=f(y)$,   with  an  absorbing  boundary  at
$x=0$.  We  show  that,  in  general, the  exponent  $\theta$  depends
continuously on the model  parameters, except when the function $f(y)$
is  an odd function,  $f(-y)=-f(y)$, when  we find  $\theta=1/4$.  Two
specific examples with odd $f(y)$ have attracted some attention in the
literature.  The  first, $f(y)=y$,  is  equivalent  to  $d^2 x/dt^2  =
\eta(t)$,  the `random  acceleration  model', for  which the  survival
probability  is known  rigorously  to decay  as  $Q(t) \sim  t^{-1/4}$
\cite{RA}.  The second  model for which we are  aware of previous work
is  the case  $f(y)=v\,{\rm sgn(y)}$.  Redner and  Krapivsky \cite{RK}
have presented qualitative arguments  that $\theta=1/4$ for this model
too,  and indeed  have conjectured  that $\theta  = 1/4$  for  all odd
functions $f(y)$.  Below  we present an argument, based  on the Sparre
Andersen theorem  \cite{SA}, that this conjecture  is correct.  First,
however,  we derive  the  result  analytically for  the  class of  odd
functions $f(y) = v\,  {\rm sgn}(y) |y|^\alpha$.  Furthermore, we will
show that  when the  amplitude $v$ takes  different values,  $v_+$ and
$v_-$, for $y>0$ and $y<0$ respectively, the exponent $\theta$ becomes
a  nontrivial function,  whose form  we determine  explicitly,  of the
exponent $\alpha$  and the ratio  $v_+/v_-$ . Our result  thus greatly
extends the  class of  first-passage problems for  which the  value of
$\theta$ is known exactly.

The starting point of the calculation is the two equations
\begin{eqnarray}
\label{langevin1}
\dot{y} & = & \eta(t)\ , \\ 
\dot{x} & = & f(y) = v_{\pm} {\rm sgn}(y)\,|y|^\alpha\ ,
\label{langevin2}
\end{eqnarray}
where the  upper (lower) sign  refers to $y>0$ ($y<0$),  dots indicate
time derivatives,  and $\eta(t)$ is  Gaussian white noise.   The final
result for $\theta$ can be written in the form
\begin{equation}
\theta            =           \frac{1}{4}-\frac{1}{2\pi\beta}\tan^{-1}
\left[\frac{v_+^\beta    -     v_-^\beta}{v_+^\beta    +    v_-^\beta}
\tan\left(\frac{\pi\beta}{2}\right)\right]
\label{result}
\end{equation}
where $\beta =  1/(2+\alpha)$.  Note that  when  $v_+=v_-$ the  result
reduces  to $\theta=1/4$,  independent of  $\alpha$, as  claimed.  The
general result is easily understood in the limits $v_- \to 0$ and $v_+
\to 0$. For  $v_- \to 0$, there is no mechanism  by which the particle
can reach the absorbing boundary at $x=0$, so $\theta =0$ as predicted
by (\ref{result}). For  $v_+ \to 0$, in order  to survive the particle
should avoid the $y<0$ region.  The probability to remain in the $y>0$
region  decays as $t^{-1/2}$  as in  the one-dimensional  random walk,
i.e.\ $\theta = 1/2$, again in accord with (\ref{result}).

We  now outline  the  derivation of  Eq.\  (\ref{result}). From  Eqs.\
(\ref{langevin1}) and (\ref{langevin2}) we can write down the backward
Fokker-Planck equation
\begin{equation}
\frac{\partial Q}{\partial t} = D\frac{\partial^2 Q}{\partial y^2} \pm
v_{\pm} (\pm y)^\alpha \frac{\partial Q}{\partial x}\ ,
\label{BFPE}
\end{equation}
where $Q(x,y,t)$  is the probability that the  particle still survives
at time $t$ given that it started at  position ($x$, $y$), and we will 
take $x \ge 0$. The absorbing boundary at $x=0$ leads to the boundary
condition  $Q(0,y,t) =  0$ for  all  $y<0$. The  initial condition  is
$Q(x,y,0)=1$ for all $x>0$.

Solving  the  full initial  value  problem  is  difficult so  we  will
specialize  to  the  late-time  scaling regime  where  $Q(x,y,t)  \sim
t^{-\theta}$.  This  approach exploits a generalization  of the method
recently   introduced  by   Burkhardt  for   $\alpha=1$   (the  random
acceleration problem) \cite{Burkhardt}.  Dimensional arguments give us
the appropriate variables for the problem.  From (\ref{langevin1}) and
(\ref{langevin2}) we  obtain $y \sim  (Dt)^{1/2}$ and $\dot{x}  \sim v
(Dt)^{\alpha/2}$    giving    $x    \sim   (v/D)y^{1/\beta}$,    where
$\beta=1/(2+\alpha)$ and  $v$ carries the dimension  of the amplitudes
$v_{\pm}$.    The   combination   $z=vy^{1/\beta}/Dx$  serves   as   a
dimensionless scaling  variable, which can also be  seen directly from
Eq.\  (\ref{BFPE}).   This  gives,   for  large  $t$,  $Q(x,t,y)  \sim
(y^2/Dt)^\theta G(z)$  where $G(z)$ is  a scaling function. This  is a
little loose, however, as we  expect different solutions for $y>0$ and
$y<0$ due  to the non-analyticity \cite{footnote} of  $f(y)$ at $y=0$.
Taking this into account, and  expressing the prefactor of the scaling
function in  terms of $x$ rather  than $y$ for  convenience, we write,
for $t \to \infty$,
\begin{equation}
Q(x,y,t)  \sim \left(\frac{x^{2\beta}}{t}\right)^\theta F_\pm\left(\pm
\frac{v_{\pm}(\pm y)^{1/\beta}}{Dx}\right)\ ,
\label{scaling}
\end{equation}
where  $F_\pm(z)$ is  the scaling  function for  $y>0$ ($+$)  or $y<0$
($-$).   The (dimensional) prefactors  (for $y>0$  and $y<0$)  in Eq.\
(\ref{scaling})   have  been  omitted   since  Eq.\   (\ref{BFPE})  is
linear. The functions $F_+(z)$ and  $F_-(z)$ are defined such that the
prefactor is the same for $y>0$ and $y<0$.

Inserting  the form  (\ref{scaling}) into  the  backward Fokker-Planck
equation  (\ref{BFPE}), we  see  immediately that  the term  $\partial
Q/\partial t$  leads to  a term of order  $t^{-(\theta +1)}$, which is
subdominant for large $t$ and  can therefore be dropped. The remaining
terms give
\begin{equation}
z F_\pm''(z) +  (1- \beta  - \beta^2 z) F_\pm'(z) +  2\beta^3 \theta
F_\pm(z)=0\ .
\end{equation}
Expressed in terms of the variable $u=\beta^2z$, this equation becomes
Kummer's   equation.   Independent   solutions   are   the   confluent
hypergeometric   functions   $M(-2\beta\theta,1-\beta,\beta^2z)$   and
$U(-2\beta\theta,1-\beta,\beta^2z)$    \cite{AS}.    The   appropriate
solutions for each  domain are selected from the  limiting behavior as
$x \to 0$.  For $y>0$, we require $Q(0,y,t) \sim (y^2/t)^\theta$ which
implies,    from    Eq.\    (\ref{scaling}),   that    $F_+(z)    \sim
z^{2\beta\theta}$      for      large      $z$.      The      function
$M(-2\beta\theta,1-\beta,\beta^2z)$ diverges  exponentially for $z \to
\infty$, so must be rejected for  $y>0$. For $y<0$, $Q$ must vanish at
$x=0+$ (because  the flow field  immediately takes the particle  on to
the    absorbing   boundary).   For    large   negative    $z$,   both
$M(-2\beta\theta,1-\beta,\beta^2z)$                                and
$U(-2\beta\theta,1-\beta,\beta^2z)$  behave  as $(-z)^{2\beta\theta}$,
so in  this regime we  retain both solutions  and fix $\theta$  by the
requirement    that   the    coefficient    of   $(-z)^{2\beta\theta}$
vanishes. Thus we write
\begin{eqnarray}
F_+(z)       &        =       &       A\,U\left(-2\beta\theta,1-\beta,
\frac{v_+\beta^2y^{1/\beta}}{Dx}\right)   \\     F_-(z)    &    =    &
B\,U\left(-2\beta\theta,1-\beta,
\frac{-v_-\beta^2(-y)^{1/\beta}}{Dx}\right)    \nonumber     \\    &&+
C\,M\left(-2\beta\theta,1-\beta,\frac{-v_-\beta^2(-y)^{1/\beta}}{Dx}
\right).
\end{eqnarray}

Relations between the  coefficients $A$, $B$, and $C$  can be obtained
by imposing the appropriate  continuity conditions at $y=0$. Requiring
$F_+(0)=F_-(0)$ gives,  using $M(a,b,0)=1$ and,  for $b<1$, $U(a,b,0)=
\pi[\sin(\pi b)\Gamma(b)\Gamma(1+a-b)]^{-1}$ \cite{AS},
\begin{equation}
C     =    (A-B)\,\frac{\pi}{\sin(\pi\beta)}\,\frac{1}{\Gamma[1-\beta]
\Gamma[\beta(1-2\theta)]},
\label{matching1}
\end{equation}
for all $\beta > 0$. The second  relation between $A$, $B$, and $C$
is  obtained by  noting that,  for  $\alpha >  -1$, Eq.\  (\ref{BFPE})
implies  that  the  first   derivative,  $\partial  Q/\partial  y$, is
continuous at $y=0$. The term linear in $y$ comes from a term of order
$z^{1-b}$  in  the  the   small-$z$  expansion  of  $U(a,b,z)$.  Since
$b=1-\beta$ here, this term is of order $z^\beta$, i.e.\ linear in $y$
since the scaling variable $z$ is proportional to $y^{1/\beta}$, as 
in Eq.\ (\ref{scaling}). Demanding that this linear term has the  same
coefficient in both regimes ($y>0$ and $y<0$) yields
\begin{equation}
Av_+^\beta = Bv_-^\beta .
\label{matching2}
\end{equation}

Finally  we  need the  asymptotic  behavior  of  $F_-(z)$ for  $z \to
-\infty$.  From the asymptotic  properties of the functions $M(a,b,z)$
and $U(a,b,z)$ \cite{AS} one readily obtains 
\begin{eqnarray}
F_-(z) &\to&
\left[B\,\frac{[\sin[\pi\beta(1-4\theta)/2]}{\sin[\pi\beta/2]} +
C\,\frac{\Gamma[1-\beta]}{\Gamma[1-\beta(1-2\theta)]}\right] \nonumber
\\ && \times (-z)^{2\beta\theta},\hspace{1cm} z \to -\infty.
\label{asymptotic}
\end{eqnarray}
Since $F_-(z)$ should vanish for $z \to -\infty$, due to the absorbing
boundary  at  $x=0$,  the  coefficient of  $(-z)^{2\beta\theta}$  must
vanish in Eq.\ (\ref{asymptotic}). This condition, combined with Eqs.\
(\ref{matching1}) and (\ref{matching2}), gives the result presented in
Eq.\ (\ref{result}).

The case  $v_+=v_-$, where  $f(y)$ is an  odd function,  is especially
simple.  Eqs.\ (\ref{matching2})  and (\ref{matching1}) give $A=B$ and
$C=0$ respectively for this case whence, from Eq.\ (\ref{asymptotic}),
one  obtains $\sin[\pi\beta(1-4\theta)/2] =  0$. The  desired solution
corresponds  to the  smallest positive  value for  $\theta$,  which is
$\theta=1/4$.   Redner and Krapivsky  \cite{RK} have  conjectured that
$\theta=1/4$ for  {\em all} odd  functions $f(y)$. A  general argument
for  this  result  can  be  made using  the  Sparre  Andersen  theorem
\cite{SA}. The theorem deals  with a one-dimensional random walk, with
an absorbing  boundary at $x=0$  and step sizes that  are independent,
identically  distributed random variables  drawn from  any continuous,
symmetric  distribution.   The  result  of  the theorem  is  that  the
probability that  the walker still survives after  $N$ steps decreases
as $N^{-1/2}$ for  large $N$. To apply this result  to our problem, we
treat  the  successive crossings  of  the  $x$-axis  by our  diffusing
particle as defining the steps of the walk, i.e.\ successive crossings
at   $x=x_n$  and  $x=x_{n+1}$   correspond  to   a  step   of  length
$x_{n+1}-x_{n}$    in   an   effective    random   walk    along   the
$x$-axis. Clearly the distribution of step sizes is continuous, and is
symmetric for  any odd $f(y)$. Since  the number of  crossings in time
$t$ scales as $N \sim  t^{1/2}$, the survival probability decays as $Q
\sim  N^{-1/2}  \sim  t^{-1/4}$  \cite{SM}.   If $v_+  \ne  v_-$,  the
distribution of step  sizes is no longer symmetric,  the conditions of
the Sparre  Andersen theorem  do not hold,  and $\theta$  is different
from $1/4$.  A superficially related  model is the Matheron-de Marsily
(MdM) model  \cite{MdM}, introduced to model the  dispersion of tracer
particles in porous rocks. In  this model, the function $f(y)$ in Eq.\
(\ref{langevin2}) is  a {\em random}  function of $y$ with  zero mean.
The MdM  model also  has $\theta=1/4$ \cite{Redner-Majumdar},  but the
Sparre Anderson  theorem cannot naively  be applied because  $f(y)$ is
not odd.

Returning to  our original model, we  note that two  special values of
$\alpha$ are of particular  interest.  The case $\alpha=0$ corresponds
to the original `windy cliff' model of Redner and Krapivsky \cite{RK},
generalized to  different drift velocities,  $v_+$ and $v_-$,  for $y$
positive and $y$ negative. If we transform to a frame of reference moving
at  speed $u$  along the  $x$-axis,  then in  the new  frame the  flow
velocities have values  $\pm v_\pm - u$, while  the absorbing boundary
has  velocity $-u$.  The  expression for  $\theta$  presented in  Eq.\
(\ref{result}) therefore applies equally  well to this class of moving
boundary problems. The second interesting value is the case $\alpha=1$.
Eliminating  $y$ from  Eqs.\ (\ref{langevin1})  and (\ref{langevin2}),
one obtains a random  acceleration process $\ddot{x} = v_\pm \eta(t)$,
with  different strength  noise forces  depending on  the sign  of the
velocity.

\begin{figure}
\narrowtext\centerline{\epsfxsize\columnwidth \epsfbox{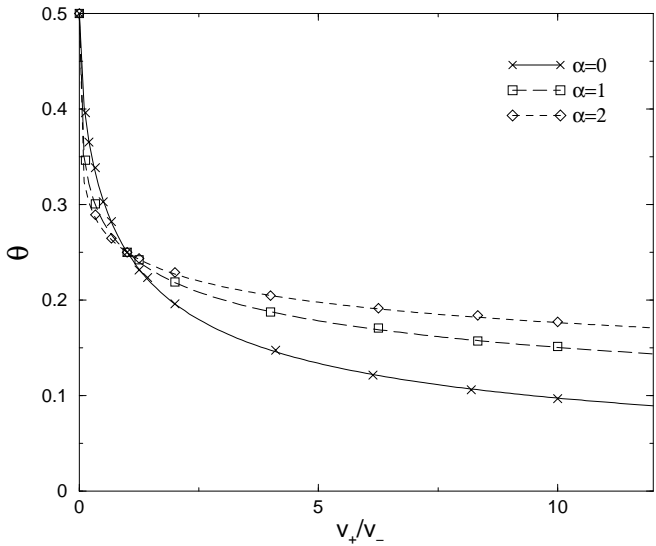}}
\caption{Simulation data for $\theta$ for a range of values of 
the exponent $\alpha$ and amplitude ratio $v_+/v_-$ in Eq.\ 
(\ref{langevin2}). The continuous curves are the predictions of 
Eq.\ (\ref{result}). The errors on the data points are smaller 
than the sizes of the symbols.}
\end{figure}

The result (\ref{result}) has been checked by computer simulations for
a  range of  $\alpha$  and  $v_+/v_-$. The  results  are presented  in
Fig.1.  Motion in  the $y$-direction  was modeled  by  a discrete-time
lattice  random  walk,  while  the  $x$ coordinate  is  treated  as  a
continuous variable. The initial  condition was $x=0$, $y=0$, with the
first step taken in the  positive $y$ direction. The simulations enter
the power-law  regime, where  $Q(t) \sim t^{-\theta}$,  quite quickly,
making it relatively straightforward to obtain an accurate estimate of
the exponent  $\theta$. The data (symbols) are  in excellent agreement
with  the predictions  (lines) in  all cases.  Notice that  all curves
and data sets cross at $\theta = 1/4$, $v_+/v_-=1$, as anticipated.

The present  exact approach is restricted  to the case  where the flow
velocity $f(y)$ has a pure  power-law form, with the {\em same} power,
for  both $y<0$  and $y>0$,  although the  amplitudes in  the  the two
regimes can be different. What can  we say about the general case? The
case  of  different powers  for  $y>0$  and  $y<0$ cannot  be  treated
analytically within  the present approach because  the assumed scaling
form,  Eq.\  (\ref{scaling}),  for  $Q(x,y,t)$  does  not  lead  to  a
consistent solution for both signs of $y$. Consider, however, the case
where the function  $f(y)$ has, for $y \to  \pm \infty$, an asymptotic
power-law  form with  the same  power for  both signs  of  $y$.  Since
trajectories  that survive  for a  long time  explore large  values of
$|y|$ \cite{RK}  we conjecture  that the exponent  $\theta$ describing
the asymptotic decay  of $Q(t)$ is given  by Eq.\ (\ref{result})
with  the exponent  $\alpha$ and  amplitude ratio  $v_+/v_-$ extracted
from the  limiting behavior  of $f(y)$ for  $y \to \pm  \infty$.  This
line  of reasoning leads  immediately to  the further  conjecture that
when  the asymptotic  behavior of  $f(y)$ for  $y \to  \pm  \infty$ is
described by  {\em different}  powers, $\alpha_\pm$, the  larger power
dominates  the $t  \to  \infty$  behavior, such  that  if $\alpha_+  >
\alpha_-$ then $\theta=0$ (i.e.\  there is a non-zero probability that
the particle escapes to infinity), while if $\alpha_+ < \alpha_-$ then
$\theta=1/2$.
  
One may also consider the  effect of diffusion in the $x$-direction as
well as  the $y$-direction.  For the  case $f(y) =  v\, {\rm sgn}(y)$,
corresponding to  $\alpha=0$, it was argued in  ref.\cite{RK} that the
diffusion in the $x$-direction is asymptotically irrelevant compared to
drift since $x_{\rm diff} \sim (Dt)^{1/2}$ whereas $x_{\rm drift} \sim
vt$.  Applying this reasoning to general $\alpha$ gives $x_{\rm drift}
\sim  t^{(2+\alpha)/2}$,  so  drift  in  the  $x$-direction  dominates
diffusion for all $\alpha >  -1$, the latter condition coinciding with
the range  of $\alpha$ for which  our analytic approach  is valid.  We
conclude that diffusion in the $x$-direction does not change the value
of $\theta$ (for $\alpha  > -1$). A more formal way to  see this is to
introduce  a  term $D_x  \partial^2Q/\partial  x^2$, corresponding  to
diffusion  in  the  $x$-direction,  on  the right-hand  side  of  Eq.\
(\ref{BFPE}).  Under  the scale  change  $y  \to  \lambda y$,  $t  \to
\lambda^2 t$, and $x \to \lambda^{1/\beta}x$, the equation retains its
form  except  for  a  rescaled  value  of  $D_x$,  given  by  $D_x'  =
\lambda^{-2(1+\alpha)}D_x$. This  shows that  $D_x$ scales to  zero on
large  length   and  time  scales  and   is  therefore  asymptotically
irrelevant.

In  summary,  we  have  obtained  exact results  for  the  persistence
exponent $\theta$ for a  class of two-dimensional stochastic processes
in  which  a  particle  diffuses   in  the  $y$-direction  and  has  a
deterministic,  $y$-dependent  drift   in  the  $x$-direction  and  an
absorbing boundary  at $x=0$.  The universal  exponent $\theta=1/4$ is
obtained  when the  drift  velocity is  an  odd function  of $y$,  but
$\theta$ has  a non-trivial  value, between zero  and $1/2$,  in other
cases.  These  results significantly expand  the (up till  now) rather
small  class  of   soluble  first-passage  problems  with  non-trivial
exponents.
 
AJB thanks Richard Blythe for useful discussions. PG's work was supported 
by EPSRC.

\end{multicols}


\begin{references}
\bibitem{footnote} The function $f(y)$ defined by Eq.\ (\ref{langevin2}) 
is nonanalytic at $y=0$ except when $\alpha$ is an odd integer and 
$v_+ = v_-$. The subsequent analysis, however, includes these cases too.
\end{references}
\end{document}